\documentclass[conference]{IEEEtran}
\IEEEoverridecommandlockouts

\usepackage{cite}
\usepackage{amsmath,amssymb,amsfonts}
\usepackage{algorithm}
\usepackage{algorithmic}
\usepackage{graphicx}
\usepackage{textcomp}
\usepackage{xcolor}
\usepackage{booktabs}
\usepackage{multirow}
\usepackage{url}

\makeatletter
\def\ps@copyrightfooter{%
    \def\@oddhead{}%
    \def\@evenhead{}%
    \def\@oddfoot{%
        \parbox{\textwidth}{%
            \raggedright
            \footnotesize
            \textcopyright~2026 IEEE. Personal use of this material is permitted.
Permission from IEEE must be obtained for all other uses, in any current or
future media, including reprinting/republishing this material for advertising
or promotional purposes, creating new collective works, for resale or
redistribution to servers or lists, or reuse of any copyrighted component of
this work in other works.
        }%
    }%
    \let\@evenfoot\@oddfoot
}
\makeatother

\begin{document}
\title{Continuous Quantum Feedback Control via Kraus-Parameterized Belief Reinforcement Learning}

\author{
\IEEEauthorblockN{Priyanshi Singh}
\IEEEauthorblockA{
Department of Computer Science and Engineering\\
SRM Institute of Science and Technology\\
Chennai, India\\
priyanshisingh.10009@gmail.com
}
\and
\IEEEauthorblockN{Krishna Bhatia}
\IEEEauthorblockA{
QuantumAI Lab\\
Fractal AI Research\\
Mumbai, India\\
krishna.bhatia@fractal.ai
}
}

\maketitle

\thispagestyle{copyrightfooter}

\begin{abstract}
Quantum feedback control requires acting on noisy continuous measurement records without direct access to the underlying quantum state. We propose Kraus-Parameterized Belief Reinforcement Learning, a pipeline in which a recurrent encoder, constrained to the Stiefel manifold, produces density-matrix estimates that are guaranteed positive-semidefinite and trace-normalized by construction, embedding quantum state geometry directly into the learning loop. A Proximal Policy Optimization (PPO) actor then maps these physically valid belief states to continuous control actions. On a simulated continuously monitored qubit, the resulting policy achieves stable feedback control, maintaining a measurement-conditioned belief fidelity of ${\sim}0.77$--$0.80$ and exhibiting substantially lower return variance than a parameter-matched LSTM-history baseline across both nominal and out-of-distribution conditions. Although gains in raw target fidelity are modest, the geometric constraint guarantees a physically valid, interpretable belief representation and yields markedly more stable control under measurement inefficiency and abrupt dynamics switches. These results indicate that physics-informed neural memory is a practical inductive bias for reliable quantum feedback control.
\end{abstract}

\begin{IEEEkeywords}
quantum feedback control, reinforcement learning, quantum filtering, Kraus
operators, belief state, partially observable control, open quantum systems
\end{IEEEkeywords}

\section{Introduction}

Quantum feedback control is inherently a problem of acting under partial
observation. In a continuously monitored qubit, the controller never directly
accesses the true density matrix; it receives a stochastic measurement record
corrupted by detector inefficiency and readout noise. The optimal feedback law
depends on the full posterior belief over the quantum state given the observed
history \cite{wiseman2010,bouten2007}. In practice, this belief is
approximated, and the quality of that approximation directly limits the
feedback performance.

Reinforcement learning (RL) offers a data-driven route to quantum feedback
without hand-crafted Lyapunov functions or linearized models
\cite{bukov2018,foesel2018,sivak2022}. A recurrent policy can
implicitly maintain a memory of past measurements and thus approximate the
belief state \cite{hausknecht2015}. However, a generic recurrent hidden state
carries no guarantee of physical consistency: its latent coordinates may not
correspond to a valid density matrix, so derived quantities such as Bloch
vector or purity lack physical meaning.

This paper asks a focused question: does an explicitly physical quantum belief
state benefit a learned feedback controller? We propose a two-stage pipeline.
First, a Kraus-constrained recurrent encoder is pretrained by supervised
regression to the simulator's true density-matrix trajectory. The Kraus
parameterization ensures that the output belief is positive semidefinite and
trace normalized at every step. Second, a PPO controller is trained on
the live quantum environment with the frozen Kraus encoder, receiving an
explicit quantum feature vector containing the estimated density matrix alongside
its purity and eigenvalue diagnostics. Because these entries encode a valid 
quantum state by construction, they provide a geometrically consistent coordinate 
system for action selection.

We compare this Kraus-belief PPO pipeline against a standard LSTM-history PPO
baseline in which a plain recurrent network replaces the belief encoder. The
comparison is run over three independently trained seeds and evaluated in
nominal, measurement-efficiency-shift, and hard-switch conditions.

\section{Problem Formulation}

We consider the task of stabilizing a continuously monitored open quantum
system (specifically a single qubit) to a target state using closed-loop
feedback. The system evolves under a control-dependent Hamiltonian and
is subjected to continuous homodyne measurement.

\subsection{System Dynamics and Measurement}
The true state of the system is described by a density matrix $\rho_t$.
The simulated stochastic master equation (SME) follows the standard
diffusive form \cite{wiseman2010}:
\begin{equation}
d\rho_t =
-i[H(u_t),\rho_t]\,dt
+\mathcal{D}[L]\rho_t\,dt
+\sqrt{\eta}\,\mathcal{H}[L]\rho_t\,dW_t,
\label{eq:sme}
\end{equation}
where $H(u_t)$ is the control-dependent Hamiltonian, $u_t$ is a scalar control
field, and $L$ is the measurement or dissipation operator. The superoperators
$\mathcal{D}$ and $\mathcal{H}$ denote the dissipative Lindblad term and
measurement innovation term, respectively. $\eta \in (0, 1]$ represents the
measurement efficiency, and $dW_t$ is a Wiener increment. 

The controller does not have access to $\rho_t$. Instead, at each discrete
time step, it receives a measurement outcome increment $dy_t$ derived from the
continuous record. 

\subsection{Task Definition}
The specific task is to drive the qubit to the pure target state $\rho^\star$
aligned with the positive $y$-axis on the Bloch sphere. The system has a fixed
drift term with angular frequency $\omega_z$ acting along the $z$-axis, and
the continuous scalar control $u_t$ acts along the $y$-axis. The continuous
control action is bounded such that $u_t \in [-4, 4]$.

The control problem is formulated as a Partially Observable Markov Decision
Process (POMDP). The RL agent seeks to maximize the expected cumulative
reward. The step reward is defined to balance target fidelity against control
effort:
\begin{equation}
r_t = F_t - \alpha \|u_t\|^2 - \beta \|u_t - u_{t-1}\|^2,
\label{eq:reward}
\end{equation}
where $F_t = \mathrm{Tr}(\rho_t\rho^\star)$ is the fidelity to the target
state, $\alpha = 0.001$ penalizes instantaneous control energy, and
$\beta = 0.0001$ penalizes action roughness to encourage smooth control
pulses. 

\section{Method}

\subsection{Data Generation and Pretraining}
To learn a valid belief state, we first generate a static dataset of simulated
quantum trajectories. We simulate 2000 trajectories of length $T=256$ steps using
a random exploratory control policy. For each step, we save the true simulated
density matrix $\rho_t$, the applied action $u_{t-1}$, and the measurement
outcome $dy_t$. 

Let $h_t = \{dy_{t-H+1}, u_{t-H+1}, \ldots, dy_t, u_{t-1}\}$ be the
recent measurement-action history of length $H=32$. The Kraus belief encoder
$f_\theta$ is tasked with processing $h_t$ to generate time-dependent Kraus operators, which iteratively update the density-matrix estimate $\hat\rho_t$. At each policy step, the encoder re-initializes $\hat\rho_0 = |0\rangle\langle 0|$ and applies the Kraus recursion across all $H=32$ steps of the current observation window.

The encoder utilizes a Long Short-Term Memory (LSTM) backbone to process the 
history $h_t$ into an unconstrained latent vector $h_t^{(L)}$. To map this standard 
neural representation to a physically valid quantum map, we utilize a specialized 
Kraus-constrained output head. First, a linear layer projects $h_t^{(L)}$ into a generic 
complex block matrix. To enforce the strict completely positive trace-preserving 
(CPTP) constraint, this matrix is orthogonalized to project it onto the Stiefel 
manifold \cite{edelman1998}, forming a valid isometry. For a single qubit, we 
utilize a Kraus rank of two. Thus, the isometry is split into two $2 \times 2$ 
operator matrices $\{K_j\}_{j=1}^2$ that perfectly satisfy the completeness relation 
by construction. The recurrent belief update is then:
\begin{equation}
\hat\rho_t = \sum_{j=1}^2 K_j\,\hat\rho_{t-1}\,K_j^\dagger,
\qquad
\sum_{j=1}^2 K_j^\dagger K_j = I.
\end{equation}
While the true continuous measurement update is non-linear, enforcing a recurrent CPTP map ensures the belief remains inside the physical state space while remaining differentiable \cite{kraus1983,choi1975}. The output $\hat\rho_t$ is unconditionally guaranteed to remain positive semidefinite ($\hat\rho_t \succeq 0$) and trace normalized ($\mathrm{Tr}(\hat\rho_t) = 1$). 

The encoder is pretrained on the offline trajectory dataset by minimizing the
Frobenius loss:
\begin{equation}
\mathcal{L}_{\text{belief}}
= \frac{1}{T}\sum_{t=1}^T \|\hat\rho_t - \rho_t\|_F^2.
\end{equation}
Once pretrained, the weights of the Kraus encoder $f_\theta$ are frozen for
the RL phase.

\subsection{Kraus-Belief PPO}
In the online RL phase, the actor interacts directly with the quantum simulator.
At step $t$, the history $h_t$ is passed through the frozen Kraus encoder to
yield $\hat\rho_t$. 

The actor receives an explicit quantum belief feature vector comprising both 
matrix and geometric information:
\begin{equation}
\phi_t = [\mathrm{vec}(\hat\rho_t),\, r_t,\, P_t,\, \lambda_{\min,t},\, h_t^{(L)},\, r^\star,\, r^\star - r_t],
\label{eq:feat}
\end{equation}
where $\mathrm{vec}(\hat\rho_t)$ contains the vectorized real and imaginary entries, $r_t$ is the estimated Bloch vector, $P_t = \mathrm{Tr}(\hat\rho_t^2)$ is the belief purity, $\lambda_{\min,t}$ is the minimum eigenvalue, $h_t^{(L)}$ is the recurrent hidden state from the LSTM, $r^\star$ is the target Bloch vector, and $r^\star - r_t$ is the Bloch error. (Because the PSD violation and trace error are exactly zero by mathematical construction, they are excluded from the network input to eliminate redundant constants). 

This composite feature vector is processed by a two-stage 128-dimensional multi-layer perceptron (MLP) architecture. It is important to emphasize that this physics-informed feature vector does not directly alter the environment dynamics; rather, it serves as a geometrically constrained observation space for the policy. The MLP maps these physically interpretable features into a policy embedding, which the PPO actor maps to the final continuous control action.

\subsection{LSTM-History Baseline}
We compare against an LSTM-history PPO baseline. This baseline passes the same
observation history $h_t$ through a standard LSTM network directly to an
actor--critic head, following the use of recurrent policies for partially
observable control \cite{hausknecht2015}. The actor also receives an
unconstrained feature vector, and no density-matrix structure is imposed.
For numerical fairness, the homodyne measurement increments passed to the LSTM
baseline are scaled by $1/\sqrt{dt}$, while the previous-action channel is left
unchanged. This conditioning gives the recurrent baseline access to measurement
fluctuations at an order-one scale without modifying the raw environment
observations or the Kraus encoder inputs. Both the Kraus-Belief model and the
LSTM baseline use hidden and feature dimensions of 128.

\begin{figure*}[t]
\centering
\includegraphics[width=\textwidth]{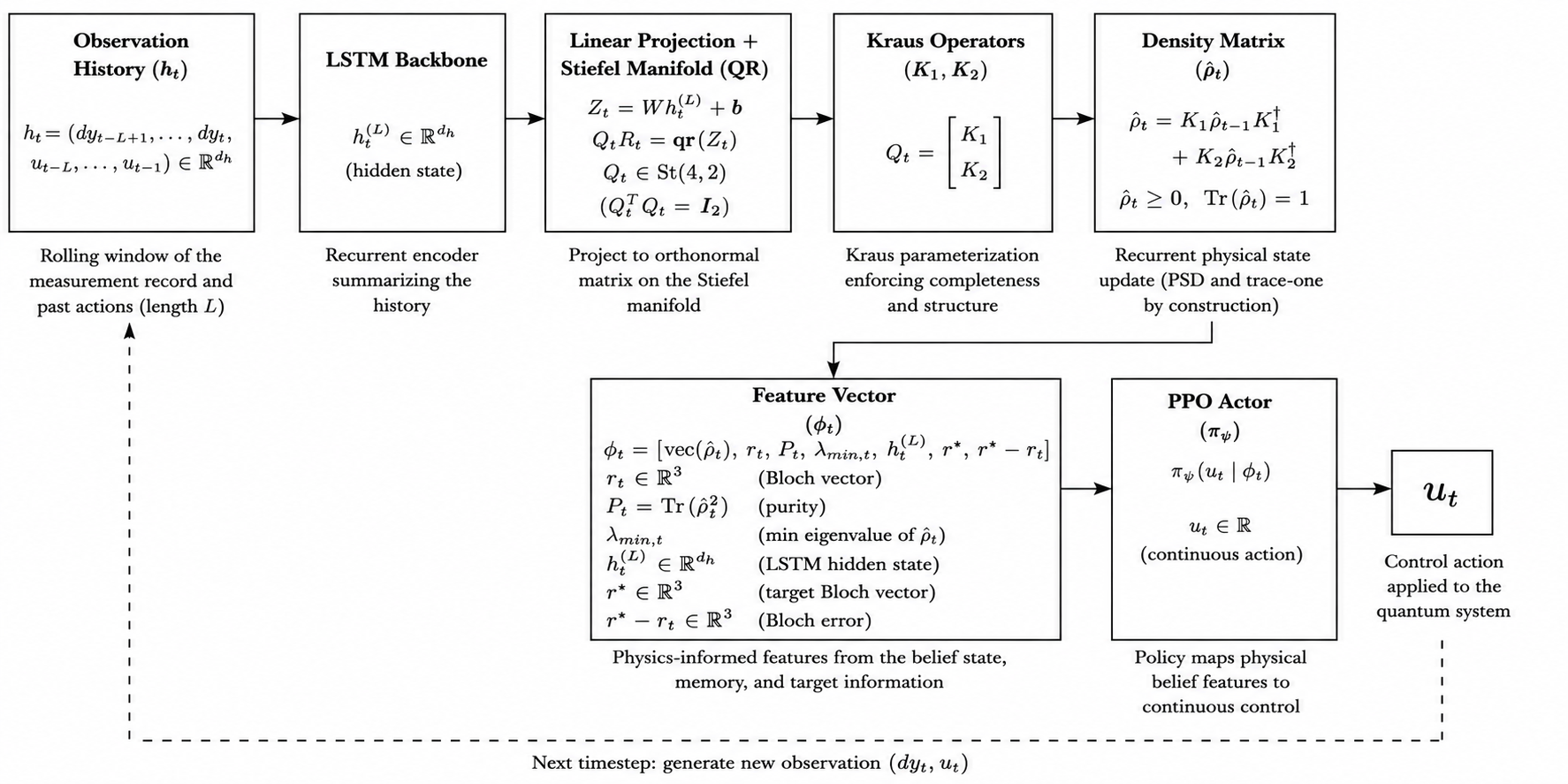}
\caption{Kraus-belief PPO pipeline. The observation history $h_t$ passes through a
frozen pretrained Kraus encoder (via a Stiefel manifold projection) to produce a 
physically valid density-matrix belief $\hat\rho_t$. Explicit physical features $\phi_t$ 
are then passed to the PPO actor. The LSTM baseline replaces this entire belief stage 
with a plain recurrent encoder.}
\label{fig:arch}
\end{figure*}

\section{Experiments}

\subsection{Setup}
During training, the environment parameters are randomized to encourage
robustness: measurement efficiency $\eta$ is drawn uniformly from $[0.6, 1.0]$, 
drift frequency $\omega_z$ is drawn from $[0.5, 2.0]$, and the damping parameter 
$\gamma$ is drawn from $[0.2, 1.0]$. Furthermore, hard-switch dynamics are enabled 
during training, with switch times drawn uniformly from $[64, 192]$ and post-switch 
$\omega_z$ drawn from $[-2.0, 2.0]$. Both Kraus-Belief PPO and the LSTM baseline 
are trained using these exact settings. Each episode is 256 steps long with a time 
step of $dt = 0.01$.

The learned, deterministic policies are evaluated under three configurations using 
identical nominal parameters ($\omega_z=1.0$, $\gamma=0.5$). All three evaluation 
conditions are run over 256 steps using deterministic policies:
\begin{itemize}
\item \textbf{Nominal}: evaluation with $\eta=1.0$ and no dynamic switch.
\item \textbf{Eta04 OOD}: evaluation with a fixed $\eta = 0.4$, pushing the 
measurements into an unseen, highly noisy regime.
\item \textbf{Hard-switch OOD}: a severe mid-episode dynamics switch where the
drift range is instantly shifted to a wider, unseen range of $[-4.0, 4.0]$.
\end{itemize}

PPO hyperparameters are identical for both methods: 300\,000 total interaction
steps, rollout length 2048, 8 update epochs per rollout, minibatch size 256,
learning rate $3{\times}10^{-4}$, discount factor $\gamma_{\mathrm{disc}} = 0.99$, GAE
parameter $\lambda = 0.95$, clipping coefficient 0.2, and an entropy coefficient 
of 0.005. Each method is trained using three independent random seeds and evaluated over
50 deterministic episodes per seed. Reported means are computed from the three
seed-level episode averages, and standard deviations are computed across those
three seed-level means.

\subsection{Metrics}
We report mean episodic return, mean target fidelity over the episode, final
terminal target fidelity, and belief fidelity (comparing the encoder's estimate to the true simulated state). 

\section{Results and Discussion}

\begin{table}[t]
\caption{Quantitative results: mean $\pm$ standard deviation across three
seed-level averages, with 50 evaluation episodes per seed. All evaluations
use 256-step episodes.}
\label{tab:results}
\centering
\scriptsize
\setlength{\tabcolsep}{3pt}
\begin{tabular}{llcccc}
\toprule
Condition & Method & Return & Mean Fid. & Final Fid. & Belief Fid. \\
\midrule
\multirow{2}{*}{Nominal}
 & Kraus & \textbf{144.4$\pm$1.8} & \textbf{0.567$\pm$0.009} & \textbf{0.538$\pm$0.044} & 0.77$\pm$0.17 \\
 & LSTM  & 131.4$\pm$23.5         & 0.516$\pm$0.094          & 0.467$\pm$0.072          & ---           \\
\midrule
\multirow{2}{*}{Eta04 OOD}
 & Kraus & \textbf{146.1$\pm$4.4} & \textbf{0.573$\pm$0.019} & \textbf{0.528$\pm$0.053} & 0.80$\pm$0.16 \\
 & LSTM  & 133.1$\pm$25.8         & 0.523$\pm$0.103          & 0.462$\pm$0.079          & ---           \\
\midrule
\multirow{2}{*}{Hard-switch}
 & Kraus & \textbf{138.4$\pm$3.6} & \textbf{0.543$\pm$0.016} & \textbf{0.525$\pm$0.007} & 0.77$\pm$0.17 \\
 & LSTM  & 132.3$\pm$14.0         & 0.520$\pm$0.057          & 0.505$\pm$0.033          & ---           \\
\bottomrule
\end{tabular}
\end{table}

Table~\ref{tab:results} reports the main quantitative comparison. For reference, a zero-control baseline (no actuation) achieves a mean fidelity of exactly $0.500 \pm 0.000$ against the $+y$ target across 50 episodes, as the uncontrolled qubit decoheres to a maximally mixed state. Several consistent patterns emerge across all three conditions.

\textbf{Return and fidelity.} Kraus-belief PPO achieves slightly higher mean
return and mean fidelity than LSTM-history PPO in every evaluated condition.
The improvement in mean return is consistent, while the reduction in return variance becomes much more pronounced under the harder out-of-distribution conditions. The mild improvement under Eta04 OOD relative to Nominal evaluation is plausible because $\eta=0.4$ remains only moderately below the training-time floor of $\eta=0.6$, rather than a severe distributional shift.

\textbf{Belief validity and state tracking.} As reported in Table~\ref{tab:results}, the Kraus encoder maintains a belief fidelity (e.g., $0.77 \pm 0.17$ nominally and $0.80 \pm 0.16$ under Eta04 OOD) relative to the true hidden state. Crucially, it guarantees positive semidefinite, trace-normalized density-matrix outputs by construction. Numerical verification confirms zero PSD violation and trace error below $10^{-7}$ in all evaluated episodes. The explicit geometric features consumed by the actor are therefore physically valid representations, ensuring the policy operates on bounded, interpretable coordinates. The LSTM baseline provides no analogous guarantee.

\textbf{Variance under distribution shift.} The standard deviation of return
across seeds is consistently smaller for Kraus-belief PPO than for LSTM in
the OOD conditions (Eta04: 4.4 vs 25.8; hard-switch: 3.6 vs 14.0). This suggests that the geometrically structured belief is associated with more stable feedback under the evaluated distribution shifts. Inspection of the learned baseline policies offers insight into this variance: across all three seeds, the unconstrained LSTM baseline tends to converge to a near-constant action (cross-episode action standard deviation $\approx 10^{-6}$), trivially minimizing the action-roughness penalty while effectively disregarding the noisy measurement record. The Kraus-belief policy instead actively modulates its bias in response to the homodyne-derived belief state at each step (action standard deviation $\approx 0.015$), maintaining genuine closed-loop correction. This supports the interpretation that the physical constraint structurally enforces observation-coupling.
\textbf{Limitations and Future Work.} The primary limitations of this study
are the evaluation on a single simulated qubit, one primary learning baseline,
and the use of three training seeds. Scaling this approach to multi-qubit
systems presents a distinct challenge: the dimension of the density matrix
grows exponentially, making a direct Kraus parameterization computationally
expensive. Future work will investigate embedding tensor-network structures
into the recurrent architecture to maintain physicality constraints in
higher-dimensional quantum spaces, as well as applying the controller to
experimental hardware.

\section{Related Work}

Quantum feedback control under continuous measurement is deeply tied to quantum filtering theory \cite{wiseman2010,bouten2007}. While traditional methods rely on explicit Lyapunov functions, data-driven approaches are increasingly necessary for systems with intractable noise models. Recent works have successfully inferred individual quantum trajectories from raw observations \cite{flurin2020} and utilized deep RL for feedback control from continuous records \cite{borah2021,porotti2020,borah2023,genois2021}. More recently, large-scale sequence models like Transformers have achieved near-unit fidelity for two-level stabilization \cite{puviani2026}. While such high-capacity models maximize raw fidelity, our contribution is fundamentally orthogonal: rather than feeding raw measurement records to unconstrained or attention-based policies, we enforce completely positive, trace-preserving (CPTP) constraints strictly on the agent's internal memory via matrix manifold optimization \cite{edelman1998}. This establishes a lightweight, strictly physics-preserving inductive bias that prioritizes exact physical interpretability at every step while preventing observation neglect.

\section{Conclusion}

We presented a Kraus-constrained belief RL pipeline for closed-loop qubit
feedback control. A pretrained Kraus encoder maps the measurement-action
history to a physically valid density-matrix belief, from which explicit
physical features are extracted for a PPO controller. Across three evaluation
conditions and three training seeds, Kraus-belief PPO achieves modestly higher
mean return and fidelity than the LSTM-history baseline. Most importantly, the
proposed pipeline avoids the open-loop collapse observed in the LSTM-history
baseline, remains responsive to the measurement record, and exhibits
substantially lower return variance in the evaluated conditions. The belief
state is positive semidefinite and trace normalized by construction, making
its features directly physical and interpretable. These results indicate that
embedding quantum-state geometry into the RL feedback loop is a promising
design choice for measurement-based quantum control.

\end{document}